# Elastic wave velocities in finitely pre-stretched soft fibers

Shiheng Zhao, Zheng Chang[*]

College of Science, China Agricultural University, Beijing 100083, China

**Abstract**

Elastic wave velocity in a soft fiber that varies depending on material constitution and axial stress level is an essential measure of mechanical signals in many technical applications. In this work, based on the small-on-large theory, we establish a model of linear elastic wave propagation in a finitely pre-stretched soft fiber. The formulas of longitudinal (Primary, P-) and transverse (Secondary, S-) wave velocities are provided and validated by numerical simulations as well as by experimental data on spider silk. The influences of material constitution, compressibility, and pre-stress on the wave propagation are investigated. We found that with increasing pre-stress, the variation of P-wave velocity highly relies on the concavity of the stress-strain curve. In contrast, an increase of S-wave velocity exhibits regardless of any constitutive model. For both P- and S-waves, the variation of the velocities is more significant in a compressible fiber than that in a nearly-incompressible one. Moreover, for minuscule pre-stress, we propose a modified formula for S-wave velocity based on the Rayleigh beam theory, which reveals the competition mechanism between "string vibration" and "beam vibration." This may provide a reliable theoretical basis for precise mechanical characterization of soft fibers and open a route for lightweight, tunable wave manipulation devices.

**Keywords**：elastic wave; wave velocity; hyperelasticity; soft fiber; pre-stress; finite deformation.

---

[*] Author to whom correspondence should be addressed. Electronic mail: changzh@cau.edu.cn (Z. Chang).

# 1. Introduction

Many soft materials are fibrous in structure. Taking advantage of such a simple and efficient "unidirectional" configuration, they are ubiquitous in nature (Gerbode et al., 2012; Jeffery L. Yarger, 2018; Saheb and Jog, 1999; Valentini et al., 2018) and possess an increasingly expanding and burgeoning range of engineering applications (Haines et al., 2014; Li et al., 2015). Whereas many miraculous soft fiber features have been revealed in different branches of physics (Gazzola et al., 2018; Huang et al., 2012; Huby et al., 2013; Lu et al., 2016; Mortimer et al., 2016), of particular interest are elastic wave responses (Beugnot and Laude, 2012; Laude et al., 2005; Schneider et al., 2016), which recently drew attention. On the one hand, this occurrence provides an effective means for in situ mechanical characterizations of soft fibers (Drodge et al., 2012; Koski et al., 2013). On the other hand, thanks to reversible material and geometrical properties accompanying finite deformation, soft fibers, as components of soft control and sensing systems (Huang et al., 2018; Miniaci et al., 2016), are attractive media for manipulating energy and information carried by elastic waves.

Soft fibers are usually subject to pre-stress in order to form a stable configuration. The propagation of elastic waves in a pre-stressed fiber is a classical problem and can be referred to in any textbook of differential equations in mathematical physics. According to the equilibrium of the infinitesimal element, the one-dimensional wave equations of P- and S-waves can be derived, and the corresponding phase velocities can be obtained as (Guenther and Lee, 1996)

$$V_P = \sqrt{\frac{E}{\rho}}, V_S = \sqrt{\frac{T}{\rho_l}}, \tag{1}$$

in which $E$ is Young's modulus, $T$ is the axial force. $\rho$ and $\rho_l$ are the volumetric and linear mass densities, respectively. However, Eq. (1) can easily be misused by overlooking the difference between pre-stressed and stretchable concepts. In case the fiber is soft (i.e., stretchable), Eq. (1) may no longer become suitable for a precise description of wave behavior due to the material nonlinearity; the distinction between the initial and current configurations resulted in this huge deformation.

Unexpectedly, the behavior of linear elastic waves propagating in a finitely pre-stretched soft fiber has not been systematically investigated, whereas the impact disturbance problem has been studied extensively (Beatty and Haddow, 1985; Drodge et al., 2012; Nowinski, 1965). However, impact-oriented research limits wave velocity to be expressed in terms of stress (in which the finite strain is inhomogeneous and flowing), which keep questions about the interplay between material properties (e.g., initial material parameters and material constitutive), structural configurations (e.g., diameter and configuration of the cross-section), and external stimuli (e.g., stress level or degree of deformation), and their contributions to the wave responses (e.g., wave velocity) remain open. Such deficiencies may lead to some improper use of the theory in many state-of-art applications. For instance, $V_{\mathrm{P}} = \sqrt{E/\rho}$ (Mortimer et al., 2014) and $V_{\mathrm{S}} = \sqrt{\mu/\rho}$ (Koski et al., 2013) (in which $\mu$ is the shear modulus and also called the second *Lamé* constant) have been used in scientific literature in characterizing the mechanical properties of pre-stretched spider silk. However, the deduced moduli based on these formulas are actually "instantaneous" moduli under tensile deformation, which can't fully reflect the mechanical properties of silk fiber.

Another concern is the S-wave propagation in a soft fiber at low pre-stress. The consensus is that the axial force governs the S-wave velocity in a soft fiber (see Eq. (1)). However, the velocity approaches zero in case the force is minute. This contradicts the fact that the S-wave exists even in a stress-free state, which is governed by a beam equation, and thus indicates the limitation of the string-based analysis. Therefore, a comprehensive understanding of the wave response at a low-pre-stress is an unmet need, which is relevant especially in applications where high-precision is demanded.

In this study, we established a model of linear elastic wave propagation in a finitely pre-stretched soft fiber based on the small-on-large theory (Norris and Parnell, 2012; Ogden, 2007). The formulas of P- and S-wave velocities expressed by the effective moduli and instantaneous mass density are provided and validated by numerical simulations as well as by experimental data on spider silk. The physical interpretation of the effective moduli is also presented to clarify their interplay with the engineering stress and elongation. The influences of material constitution, compressibility, and pre-stress on the wave propagation are investigated. We found that with an increasing pre-stress, the variation of P-wave velocity heavily relies on the concavity of the stress-strain curve. In contrast, the increasing S-wave velocity exhibits regardless of any constitutive model. For both P- and S-waves, the variation of the velocities is more significant in a compressible fiber than that in an incompressible one. Moreover, for minuscule pre-stress, we propose a modified formula for S-wave velocity based on the Rayleigh beam theory, which clearly reveals the competition mechanism between "string vibration" and "beam vibration."

The paper is organized as follows: in Sec. 2, the small-on-large theory, which describes linear wave motion propagation in a finitely deformed hyperelastic material, is briefly reviewed. In Sec. 3, the governing equations of elastic waves in pre-stressed soft fibers are derived, and the analytical and numerical procedures to obtain the wave velocities are provided. In Sec. 4, two examples are illustrated to compare with numerical simulations and experimental data. Moreover, the influences of the constitutive model, compressibility, and pre-stress on wave propagation are discussed. Finally, a discussion on our results and on avenues for future work is provided in Sec. 5.

## 2. Theoretical background: small-on-large theory

The small-on-large theory is employed to determine how small-amplitude elastic waves propagate in a finitely pre-stretched fiber. In the theory, the static finite deformation and wave motion are decoupled based on a linearization of the static finite-deformation state.

### 2.1. Static finite deformation

Consider a hyperelastic solid with constitutive behavior characterized by the strain energy function (SEF) $W$. The "finite deformation" $U_i = x_i(X_j)$ (or more rigorously, the current position $x_i$ of a material particle originally located at $X_j$ after deformation) is defined in terms of rectangular *Cartesian* components. $x_i$ and $X_j$ denote the current (deformed) and reference (undeformed) configurations, respectively. The large deformation (free of body force) follows the static equilibrium equation of

$$S_{ij,i} = 0, \tag{2}$$

where

$$S_{ij} = \frac{\partial W}{\partial F_{ji}} \tag{3}$$

denotes the nominal (or first *Piola-Kirchhoff*) stress tensor, $F_{ij} = \partial x_i / \partial X_j$ is the deformation gradient. By substituting the relation

$$A_{ijkl} = \frac{\partial S_{ij}}{\partial F_{lk}} = \frac{\partial^2 W}{\partial F_{ji} \partial F_{lk}}, \tag{4}$$

into Eq. (2), the equilibrium equation is then written (Ogden, 1997)

$$(A_{ijkl} U_{l,k})_{,i} = 0. \tag{5}$$

$A_{ijkl}$ is non-linearly determined by $F_{ij}$. Therefore, Eq. (5) refers to a set of quasi-linear partial differential equations of the second order for $U_i$. In terms of a particular $U_i$ (or equivalently $F_{ij}$), the uniquely determined $A_{ijkl}$ is termed as the instantaneous elastic tensor expressed in the reference configuration. Both $A_{ijkl}$ and $U_i$ are utilized to analyze incremental wave motion.

## 2.2. Incremental linear wave motion

The incremental linear wave motion $u_i$ superimposed onto the finite deformation $U_i$ complies with (Ogden, 1997)

$$(\sigma_{i'j})_{,i'} = \rho' \ddot{u}_j, \tag{6}$$

where

$$\sigma_{i'j} = A'_{i'jk'l} u_{l,k'} \tag{7}$$

and $\rho'$ are the true (or *Cauchy*) stress and the mass density expressed in the current configuration. In Eq. (6) and (7), $A'_{i'jk'l}$ and $\rho'$ can be obtained by pushing forward on $A_{ijkl}$ and $\rho$ (expressed in the reference configuration), respectively, i.e., (Ogden, 2007)

$$A'_{i'jk'l} = J^{-1} F_{i'i} F_{k'k} A_{ijkl}, \quad \rho' = J^{-1} \rho, \tag{8}$$

where $J = \det(F_{ij})$ is the volumetric ratio. Due to this forward push, the effective elastic tensor in the current configuration loses small symmetry, which implies that the material can be regarded as a Cosserat-like material (Eringen, 2012). In contrast with Eq. (5), Eq. (6) are a set of linear partial differential equations of the second order for $u_i$.

## 3. Elastic wave velocities in finitely pre-stretched soft fibers
### 3.1. Model description

We start by focusing on a piece of hyperelastic fiber with infinite length, which is shown in Fig. 1. For simplicity, the two-dimensional (2D) plane-strain assumption is applied to analyze the three-dimensional (3D) problem. According to the assumption, the out-of-plane principal stretch of the finite deformation $F_{33} \equiv 1$, thus the SEF can simply degenerate into a 2D form. However, in considering the fiber's geometry, as its width (in out-of-plane direction) is significantly shorter than its length, plane-stress is more suitable. Therefore, a substitution for the initial material parameters is applied to obtain the plane-stress solution from a plane-strain problem, i.e., (Ugural and Fenster, 2003)

$$\lambda_{2D} = \frac{2\lambda_{3D}\mu_{3D}}{\lambda_{3D} + 2\mu_{3D}}, \qquad (9)$$
$$\mu_{2D} = \mu_{3D},$$

in which $\lambda$ and $\mu$ are the initial *Lamé* constants. The subscript "3D" denotes the material parameter in a 3D (or equivalently 2D plane-stress) problem, while "2D" represents the transformed material parameter applied in the corresponding 2D plane-strain problem.

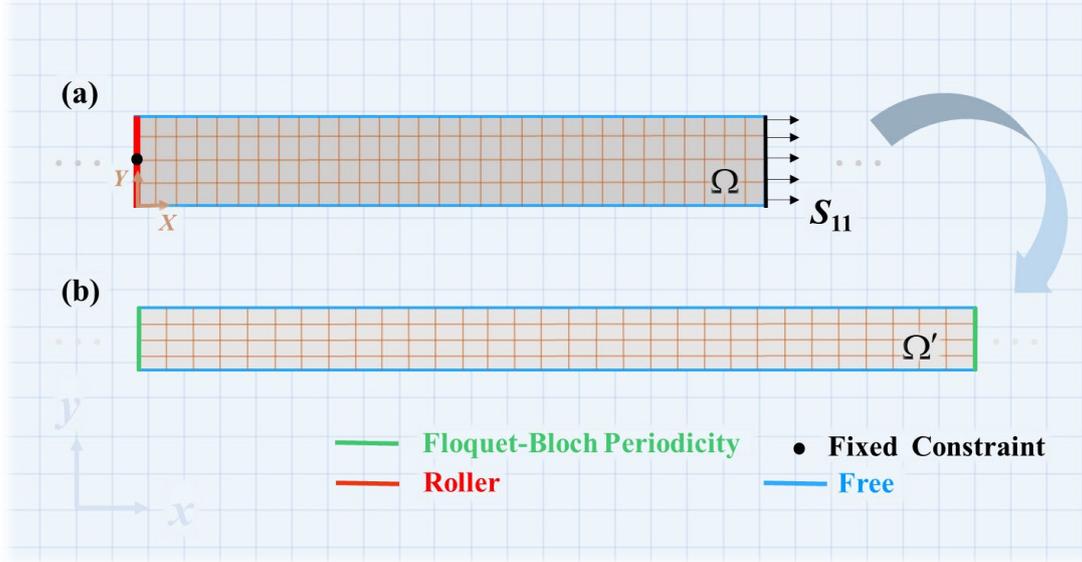

FIG. 1. Schematic diagram of a piece of 2D hyperelastic fiber with infinite length. (a) and (b) demonstrate the configurations before ($\Omega$) and after ($\Omega'$) the finite tensile stretch. The boundary conditions applied in the analyses of finite deformation and linear wave propagation are illustrated in (a) and (b), respectively. The deformed configuration ($\Omega'$) is utilized as a primitive cell in the wave analysis.

As is illustrated in Fig. 1, the quasi-static axial tensile stress (or deformation) cause the fiber to distort from the reference configuration ($\Omega$) to the current one ($\Omega'$). Based on the small-on-large theory, the stretched fiber can be regarded as a stress-free Cosserat media. Taking advantage of the governing equation (Eq. (6)) and specific boundary conditions of fiber structure (see Fig. 1(b)), we can obtain the governing equation of the P- and S-waves, together with the corresponding wave velocities.

**3.2. Governing equations and phase velocities of P- and S-waves**

For a finitely pre-stretched soft fiber, we let the deformation gradient take a uniaxial form, i.e.

$$\mathbf{F} = \begin{bmatrix} F_{11} & 0 \\ 0 & F_{22} \end{bmatrix}, \tag{10}$$

hence $A'_{i'jk'l}$ can be expressed

$$A'_{i'jk'l} = \begin{bmatrix} \begin{bmatrix} A'_{1111} & A'_{1122} \\ A'_{2211} & A'_{2222} \end{bmatrix} & 0 \\ 0 & \begin{bmatrix} A'_{1212} & A'_{1221} \\ A'_{2112} & A'_{2121} \end{bmatrix} \end{bmatrix}, \quad (11)$$

in which only upper left and lower right sub-blocks are non-zero. Such representation implies that the horizontal and vertical particle displacements in the fiber are decoupled.

Since the upper and lower boundaries of the fiber are free of normal and shear stresses

$$\sigma_{22} = 0, \sigma_{21} = 0, \quad (12)$$

from Eq. (7) we obtain

$$\sigma_{11} = \left( A'_{1111} - \frac{A'_{1122} A'_{2211}}{A_{2222}} \right) u_{1,1}, \quad (13)$$

$$\sigma_{12} = \left( A'_{1212} - \frac{A'_{1221} A'_{2112}}{A_{2121}} \right) u_{2,1}. \quad (14)$$

In this fashion, we can define the effective Young's modulus and effective shear modulus, i.e.,

$$\tilde{E} = A'_{1111} - \frac{A'_{1122} A'_{2211}}{A'_{2222}}, \quad (15)$$

$$\tilde{G} = A'_{1212} - \frac{A'_{1221} A'_{2112}}{A'_{2121}}, \quad (16)$$

to measure the stiffness of a stretched fiber to resist incremental wave motion in P- and S-modes, respectively.

From Eq. (6), (7) and (11), we can rewrite the governing equations of P- and S-waves

$$\sigma_{11,1} + \sigma_{21,2} = \rho' \ddot{u}_1, \quad (17)$$

and

$$\sigma_{12,1} + \sigma_{22,2} = \rho' \ddot{u}_2. \quad (18)$$

After substitution of the boundary conditions Eq. (12), and in considering both the material distribution and the pre-deformation are homogeneous, Eq. (17) and (18) can

be rewritten

$$\tilde{E}u_{1,11} = \rho' \ddot{u}_1, \tag{19}$$

and

$$\tilde{G}u_{2,11} = \rho' \ddot{u}_2. \tag{20}$$

Eq. (19) and (20) are classical one-dimensional wave equations, and the corresponding velocities of P- and S-waves ($V_P$ and $V_S$) can be obtained as

$$V_P = \sqrt{\frac{\tilde{E}}{\rho'}}, \tag{21}$$

$$V_S = \sqrt{\frac{\tilde{G}}{\rho'}}. \tag{22}$$

Similarly, the effective Young's modulus and shear modulus for a 3D problem can also be derived as

$$\tilde{E} = \frac{A'_{1111}(A'_{2222}A'_{3333} - A'_{2233}A'_{3322})}{A'_{2222}A'_{3333} - A'_{3322}A'_{2233}} + \frac{A'_{1122}(A'_{2233}A'_{3311} - A'_{2211}A'_{3333})}{A'_{2222}A'_{3333} - A'_{3322}A'_{2233}} + \frac{A'_{1133}(A'_{2211}A'_{3322} - A'_{2222}A'_{3311})}{A'_{2222}A'_{3333} - A'_{3322}A'_{2233}}, \tag{23}$$

$$\tilde{G} = A'_{1212} - \frac{A'_{1221}A'_{2112}}{A'_{2121}} = A'_{1313} - \frac{A'_{1331}A'_{3113}}{A'_{3131}}. \tag{24}$$

Correspondingly, the velocities of the P- ($V_P$) and S-waves ($V_{SH}$ and $V_{SV}$ in 3D problem) have the same form as Eq. (21) and (22), respectively.

### 3.3. Physical interpretation of the effective moduli $\tilde{E}$ and $\tilde{G}$

Through some derivation, we can prove (see Appendix) that the effective Young's modulus $\tilde{E}$ equals to a forward push of the tangent modulus $\hat{E} = dS_{11}/dF_{11}$, i.e.,

$$\tilde{E} = \frac{F_{11}^2}{J}\hat{E} = \frac{F_{11}^2}{J}\frac{dS_{11}}{dF_{11}}. \tag{25}$$

Meanwhile, we can also prove (see Appendix) that the effective shear modulus $\tilde{G}$ is equivalent to the axial Cauchy stress $\sigma_{11}$, i.e.,

$$\tilde{G} = \sigma_{11} = \frac{F_{11}S_{11}}{J}. \tag{26}$$

In this fashion, Eq. (21) and (22) can be rewritten from the perspective of stress as

$$V_\mathrm{P} = \sqrt{\frac{F_{11}^2}{\rho}\frac{\mathrm{d}S_{11}}{\mathrm{d}F_{11}}}, V_\mathrm{S} = \sqrt{\frac{F_{11}S_{11}}{\rho}}. \tag{27}$$

Compared with Eq. (21) and (22), such alternative representation establishes a bridge between wave velocity, stress, and finite deformation, which is conducive to its application in experimental measurement. In addition, Eq. (27) also suggests that elongation, apart from engineering stress, tangent modulus, and initial mass density, is another essential quantity for predicting wave velocities in soft fibers. It is also worthy to note that Eq. (27) is consistent with a forward push of the Lagrangian wave speeds of longitudinal and transverse shocks (Nowinski, 1965).

**3.4. Numerical modeling**

To validate the above theoretical result, numerical simulations have been performed by a finite-element-method (FEM)-based two-step model, using the software COMSOL Multiphysics.

In the first step, the finite deformation of the hyperelastic material is calculated by using the module of solid mechanics. For the model described in Sec. 3.1, we consider a hyperelastic domain $\Omega$ (Fig. 1(a)), in which the distortion is governed by Eq. (5). With the upper and lower boundaries of $\Omega$ free of constraint, a roller is imposed on the left boundary to constrain the longitudinal displacement without affecting lateral deformation. The midpoint of the boundary is fixed to prevent rigid body translation. In this fashion, $\Omega$ can be finitely deformed to $\Omega'$ (Fig. 1(b)) by applying either force or displacement to the right boundary.

In the second step, the band structure of the soft fiber is determined by using the module of weak-form PDE. With the deformation gradient obtained in the first step being imported, we take $\Omega'$ as a primitive cell, in which the incremental wave propagation is governed by Eq. (6). *Floquet-Bloch* periodicity is imposed at the left and right boundaries, while the upper and lower boundaries are free. Through an eigenvalue analysis for $kL \in [0, \pi]$ ($k$ is the wavenumber, $L$ is the length of the primitive cell), the band structure, i.e. the dispersion relation of elastic waves, can be computed, which allows for calculation of the phase velocities of P- and S-modes, i.e.,

$$V = \frac{\omega}{k}. \qquad (28)$$

## 4. Results

In what follows, two examples are provided to verify the theory and numerical methods. The former considers the propagation of elastic waves in a compressible rubber cord under finite stretching. In this regard, the wave velocities obtained from Eq. (21) and (22) are compared with those obtained from numerical simulations. The latter focuses on the elastic waves in spider silk, where we compared the analytical result with experimental data. In this example, the velocities obtained from a 3D finite element model are also presented to address the influence of cross-section configuration on wave velocities.

### 4.1. Elastic wave velocities in a finitely pre-stretched rubber cord

Consider a rubber cord with a width of $0.01\,\text{m}$. The rubbery material obeys the neo-Hookean SEF, of which the 2D form can be written as (Chen et al., 2017; Ogden, 1997)

$$W_{\text{NH1}} = \frac{\lambda_{2D}}{2}(J-1)^2 - \mu_{2D}\ln J + \frac{\mu_{2D}}{2}(I_1 - 2), \qquad (29)$$

where $I_1 = \text{tr}(B_{ij})$ is the first invariant of the right *Cauchy-Green* tensor $B_{ij} = F_{si}F_{sj}$, According to Eq. (9), the initial *Lamé* constants $\lambda_{3D} = 4.32\,\text{MPa}$ and $\mu_{3D} = 1.08\,\text{MPa}$ (which correspond to a compressible version of PSM-4 (Bertoldi and Boyce, 2008)) are transformed into $\lambda_{2D} = 1.44\,\text{MPa}$ and $\mu_{2D} = 1.08\,\text{MPa}$, respectively. The mass density is $\rho = 1000\,\text{kg/m}^3$. In this fashion, the effective Young's modulus and shear modulus of the finitely pre-stretched rubber cord are (Chang et al., 2015)

$$\tilde{E} = \tilde{\lambda} + \tilde{\mu} + \tilde{\mu}_1 - \frac{\tilde{\lambda}^2}{\tilde{\lambda} + \tilde{\mu} + \tilde{\mu}_2},$$
$$\tilde{G} = \tilde{\mu}_1 - \frac{\tilde{\mu}^2}{\tilde{\mu}_2},$$
(30)

in which

$$\tilde{\lambda} = \lambda_{2D}(2J-1),$$
$$\tilde{\mu} = \lambda_{2D}(1-J) + J^{-1}\mu_{2D},$$
$$\tilde{\mu}_1 = J^{-1}\mu_{2D}F_{1s}F_{1s},$$
$$\tilde{\mu}_2 = J^{-1}\mu_{2D}F_{2s}F_{2s}.$$
(31)

The P- and S-wave velocities for various tensile stretch can thus be obtained by Eq. (21) and (22), as the corresponding linear dispersion relations being illustrated in Fig. 2. All the dispersion curves, except for the S-mode at a stress-free state, show an excellent agreement with numerical results in which we take a piece of the cord with an initial length of $0.1\,\text{m}$ as a primitive cell. When the cord is free of tensile stress, the numerical result displays the S-wave as dispersive, and the velocity is non-zero, which indicates a failure of our analytical model in such a condition. This issue will be addressed in detail in Sec. 4.3.3.

Fig. 2 also demonstrates the convergence of the $V_P$ and $V_S$ in the rubber cord. With nominal stress $S_{11}$ increasing from 0 to $2\,\text{MPa}$, $V_P$ increases from $54.9\,\text{m/s}$ to $75.6\,\text{m/s}$ (the decrease of the slopes of the P-modes is due to the normalization of the abscissa based on $kL$), while $V_S$ increases from approximately zero to $64.1\,\text{m/s}$. It can be analytically predicted that the wave velocities will be identical to each other when $\tilde{E} = \tilde{G}$, which gives a condition between material parameters and finite deformations for the rubber cord with the $W_{\text{NH1}}$ SEF

$$\frac{\lambda_{2D}}{\mu_{2D}} = \frac{(J^2 + F_{11}^{\ 2})}{F_{11}^{\ 2}J(J-1)}.$$
(32)

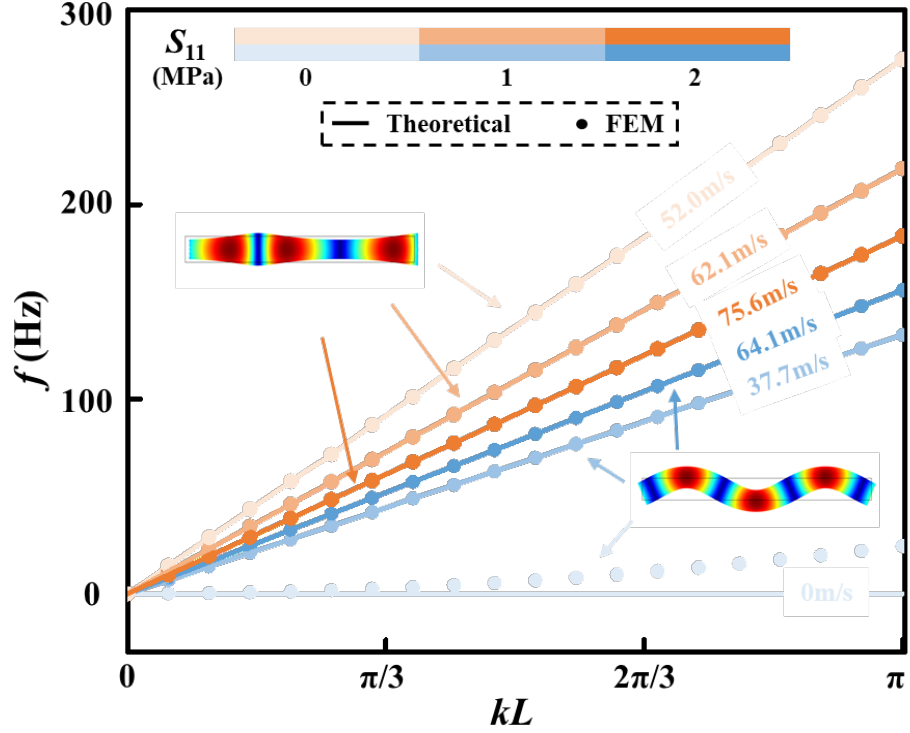

FIG. 2. Dispersion curves of P- and S-modes in pre-stretched soft fiber at various levels of nominal stress as obtained from numerical (FEM) simulations (marker) or predicted analytically (solid line). The predicted wave velocities are the slopes of the dispersion curves.

## 4.2. Elastic wave velocities in a finitely pre-stretched spider silk

Consider a *Nephila* minor ampullate spider silk whose diameter (or width in the 2D model) is $3.38 \times 10^{-6}$ m (Mortimer et al., 2014). The initial material parameters are determined from literature (Blackledge and Hayashi, 2006) as $E = 10.5$ GPa and $\rho = 1325$ kg/m$^3$. The *Poisson*'s ratio is assumed to be $\upsilon = 0.49$, as the silk is nearly incompressible. Therefore, the 2D *Lamé* constants are $\lambda_{2D} = 6.77$ GPa and $\mu_{2D} = 3.52$ GPa, respectively. For simplicity, we also assume $W_{NH1}$ characterizes the constitutive behavior of the spider silk. As the difference among the wave velocities for different SEFs is not significant (see Sec. 4.2) at a low-stress level. In this sense, $W_{NH1}$ is sufficient to capture the wave velocities within the range of tensile stress ($\leq 500$ MPa) adopted in the experimental measurement [22]. For a higher stress level, although the neo-Hookean model can't represent the spider silk, it is still enough to reveal a significant difference between our theory and formulas previously used in the literature.

Fig. 3 shows $V_P$ and $V_S$ in the pre-stretched spider silk versus nominal stress $S_{11}$ (or elongation $F_{11}$) given by the present 2D model (in Eq. (21) and (22)) and the experimental data (Mortimer et al., 2014). The S-wave velocity yields an agreement between the theoretical and experimental results. For the P-wave counterpart, the theoretical result is at the same magnitude as the experimental data, as well as the same upward trend with increasing stress. However, experimentally measured velocity proceeds at a sharper increase than that of the theoretical prediction. According to Eq. (27), $V_P$ mainly depends on the fiber's tangent modulus. The rapid increase in the velocity requires a drastic rise in the modulus. As a sudden increase in modulus hasn't been observed in any existing tensile experiments (Drodge et al., 2012; Guo et al., 2018) for spider silks, we hypothesize that the P-wave velocity may be overestimated in the experimental measurement.

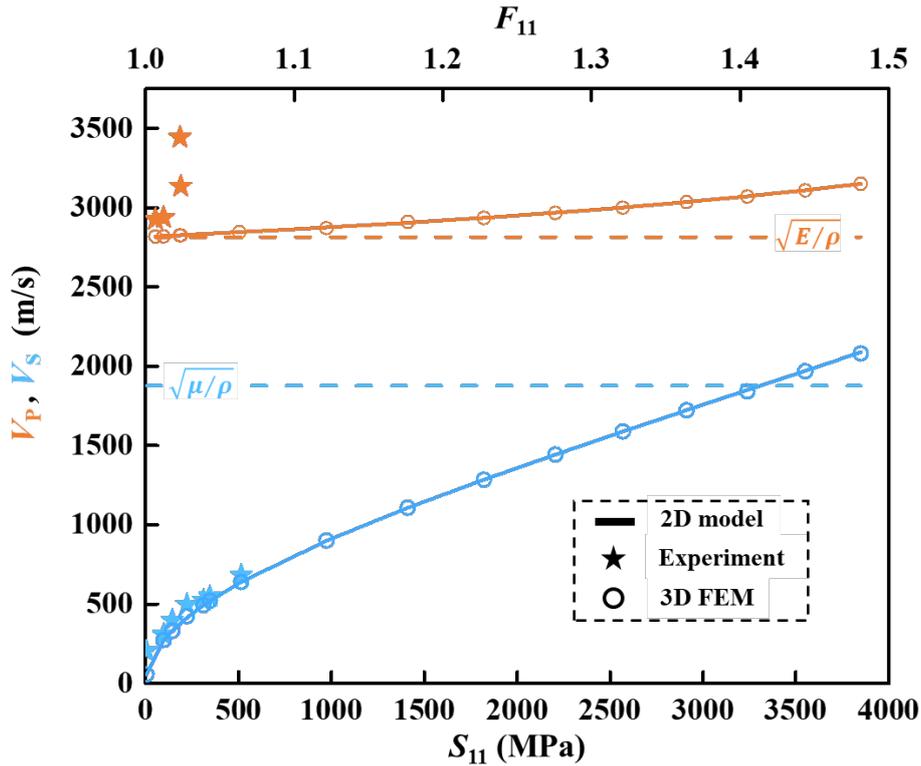

FIG. 3. Velocities of P- and S-waves in pre-stretched spider silk versus nominal stress $S_{11}$ (or elongation $F_{11}$) given by the present 2D model (in Eq. (21) and (22)), experimental data (Mortimer et al., 2014), and 3D FEM simulation. Velocities obtained from the formulas of $V_P = \sqrt{E/\rho}$ and $V_S = \sqrt{\mu/\rho}$ are also provided for comparison.

3D FEM simulations are also provided to investigate the influence of cross-sectional shape on wave velocities. Given the cross-sectional area of $1\times10^{-4}$ m$^2$, silks with circular (see Fig. 3) and square (not shown) cross-sections are considered. It shows that the cross-sectional configuration has a negligible effect on wave propagation, so that the 3D results for both cross-sections well confirm that of the 2D model.

Moreover, two formulas for the wave velocities utilized previously, i.e. $V_\mathrm{P} = \sqrt{E/\rho}$ (Mortimer et al., 2014) and $V_\mathrm{S} = \sqrt{\mu/\rho}$ (Koski et al., 2013), are also drawn in Fig. 3 as a comparison. They predict constant velocities (independent of stress), and are represented as two horizontal lines in the figure. At a low-stress level ($\leq 500$ MPa), $V_\mathrm{P} = \sqrt{E/\rho}$ (2815 m/s) shows a reasonable agreement with the theoretical result, as in this range, the elongation $F_{11}$ of the silk is relatively small and $\mathrm{d}S_{11}/\mathrm{d}F_{11} \approx E$ (see Eq. (25)). However, the discrepancy between this formula and present model will gradually increase to 11.9% (3150 m/s) as the stress increases to 4000 MPa. By contrast, $V_\mathrm{S} = \sqrt{\mu/\rho}$ fails in describing the S-wave velocity. This is because, in a spider silk subjected to tensile stretching, the transmission of shear wave motion between particles is governed by the stress, $\sigma_{11}$ (see Eq. (26)), while not the shear modulus of the bulk material, $\mu$. In other words, $\sqrt{\mu/\rho}$ is the S-wave velocity in a stress-free bulk material, whereas not suitable for a soft fibrous structure under finite extension. Regarding spider silk's stress level ($\sim 10^2$ MPa) on an orb-web (Wirth and Barth, 1992), $V_\mathrm{S} = \sqrt{\mu/\rho}$ exhibits a huge variance from the present model. Therefore, the mechanical characterization based on such a formula will overestimate the shear modulus of the silk.

**4.3. Influences of strain energy function, compressibility, and pre-stress**

In this sub-section, the effects of material constitution, compressibility, and pre-stress on the wave velocities are presented based on the example of a rubber cord proposed in Sec. 4.1.

**4.3.1. Strain energy function (SEF)**

Two types of compressible SEFs are considered. Type-I express the SEF as a function of the invariants of the right *Cauchy-Green* tensor. In addition to $W_{NH1}$ (Eq. (29)), *Gent* SEF can be expressed as (Chen et al., 2017; Gent, 1996)

$$W_{GE1} = -\frac{\mu_{2D}J_m}{2}\ln\left(1-\frac{I_1-2}{J_m}\right) - \mu_{2D}\ln J + \left(\frac{\lambda_{2D}}{2} - \frac{\mu_{2D}}{J_m}\right)(J-1)^2, \quad (33)$$

where $J_m$ is a material constant related to the strain saturation of the material. Note that $W_{GE1}$ degenerates to $W_{NH1}$ when $J_m \to \infty$. In type-II, the *Cauchy-Green* deformation tensor is decomposed by multiplication, and the SEF consists of both a deviatoric and hydrostatic term. In this sense, the corresponding neo-Hookean and *Gent* SEFs can be written as (Chen et al., 2017)

$$W_{NH2} = \frac{\mu_{2D}}{2}(\overline{I}_1 - 2) + \frac{\kappa}{2}(\ln J)^2, \quad (34)$$

and

$$W_{GE2} = -\frac{\mu_{2D}J_m}{2}\ln\left(1-\frac{\overline{I}_1-2}{J_m}\right) + \frac{\kappa}{2}(J-1)^2, \quad (35)$$

where $\overline{I}_1 = J^{-1}I_1$, and $\kappa = \lambda_{2D} + \mu_{2D}$ is the 2D bulk modulus.

Given the initial material parameters $\lambda_{2D}$ and $\mu_{2D}$ together with the geometric model the same as those used in Sec. 4.1, Fig. 4 shows the variation of wave velocities with elongation obtained by our theory for different SEFs (with the light gray area represents range in which $V_S$ is dispersive). It demonstrates for both P- and S-waves that the velocities for type-I SEFs are higher than those of type-II. With increasing elongation, unlike the uniform increase of $V_S$, different SEFs exhibit significant difference in the monotonicity of $V_P$. Particularly, $V_P$ increases monotonically for $W_{NH1}$ and $W_{GE1}$, decreases for $W_{NH2}$, whereas the first decreases and then increases for $W_{GE2}$. It also shows that the variation of $V_P$ for any SEF is consistent with the variation of the concavity of its stress-strain curves (see inset of Fig. 4). This result also reveals that the convergence of $V_P$ and $V_S$ mentioned in Sec. 4.1 is not universal, and has a strong dependence on the material model.

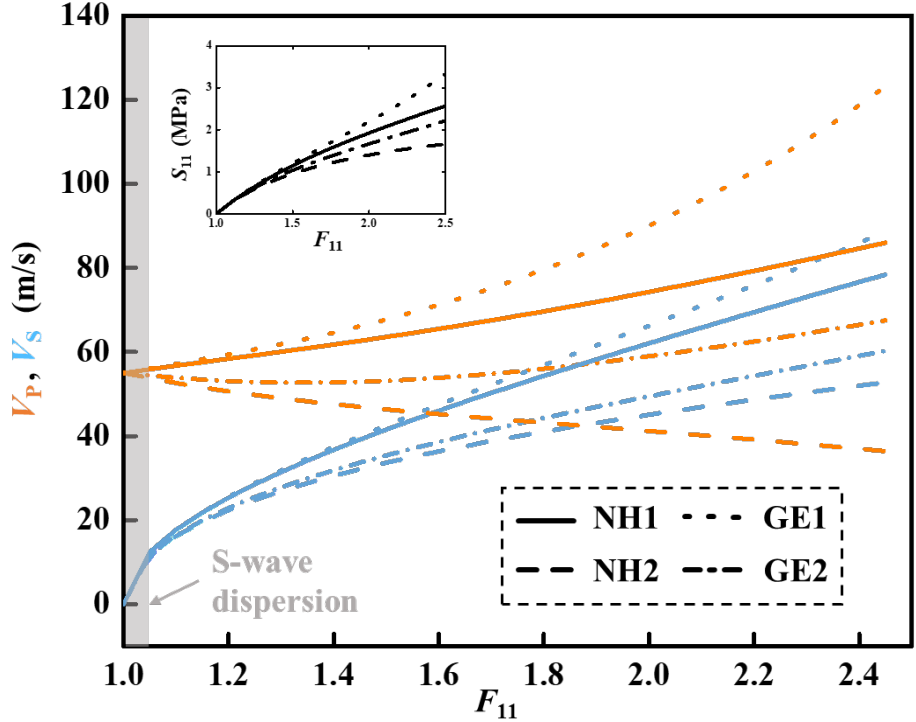

FIG. 4. Velocities of P- and S-waves in pre-stretched soft fiber versus elongation $F_{11}$ for various SEFs. The inset shows the corresponding stress $(S_{11})$-elongation $(F_{11})$ curves. The light grey area demonstrates the region in which the S-wave is dispersive.

**4.3.2. Compressibility**

Given the SEF, the first *Lamé* constant $\lambda_{3D}$, and the geometric model the same as those used in Sec. 4.1, the influence of the compressibility on the wave velocities is investigated by varying the shear modulus $\mu_{3D}$ from $\lambda_{3D}/49$ to $\lambda_{3D}$. Correspondingly, *Poisson*'s ratio decreased from $\upsilon = 0.49$ (nearly-incompressible) to $0.25$.

Fig. 5(a) and (b) displays $V_P$ and $V_S$ for various compressibility (*Poisson*'s ratio) and deformation, respectively. For both P- and S-waves, the velocities are relatively low for a nearly-incompressible fibre, and the velocities increase as the compressibility increases. Meanwhile, the influence of elongation on the increment of the velocities become more significant with increasing compressibility. It is worth noting that regardless of compressibility, the increasing rate in the velocities that result from the change in elongation is approximately identical. For instance, as $F_{11}$ increases from 1.2 to 1.8, $V_P$ increases by 23.5% (from 17.1 m/s to 21.0 m/s), while $V_S$ increases by 125.2% (from 2.1 m/s to 4.73 m/s), when $\upsilon = 0.49$. A similar growth ratio can be observed when $\upsilon = 0.3$, i.e., 23.8% (from 93.9 m/s to 116.3 m/s) for $V_P$ and 120.3% (from 36.5 m/s to 80.4 m/s) for $V_S$. Fig. 5 also reveals that within the range of elongation and *Poisson*'s ratio that we consider, the two quantities have an equivalent influence on $V_S$. In contrast, the effect of *Poisson*'s ratio on the P-wave is more significant than that of elongation.

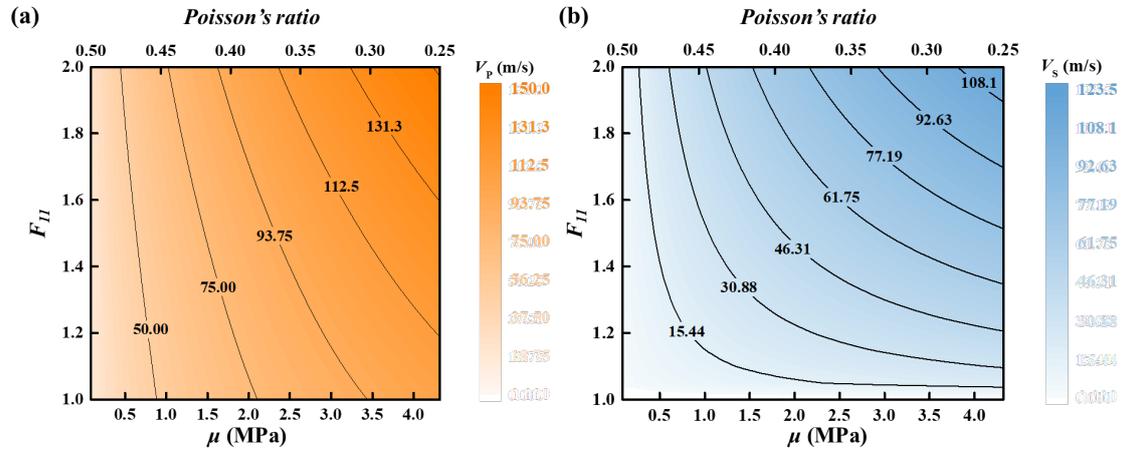

FIG. 5. Velocities of P- (a) and S- (b) waves in pre-stretched soft fiber versus material compressibility and elongation $F_{11}$. The compressibility varies as the second *Lamé* constant $\mu_{3D}$ increases from 0.0082 MPa ($\lambda_{3D}/49$) to 4.32 MPa ($\lambda_{3D}$) when the first *Lamé* constant $\lambda_{3D}$=4.32 MPa. Correspondingly, *Poisson*'s ratio decreases from 0.49 to 0.25. The solid black curves represent the contours of the velocities.

### 4.3.3. S-wave velocity at an extreme low pre-stress

The previous narrative perceives that Eq. (22) fails to evaluate the S-wave velocity in a soft fiber at a stress-free state. In the absence of axial force, $\tilde{G}$ equals zero. In this circumstance, the transverse wave propagates in the soft fiber following the "beam vibration" mechanism, while not with the "string vibration". The numerical result reported in Fig. 2 confirms this hypothesis in which the dispersive nature of "beam vibration" is demonstrated. Following this idea, we expand the *Rayleigh* beam theory (Rayleigh, 1896) and construct a beam vibration model for a finitely stretched soft fiber. In the current configuration, the beam equation is written as

$$\rho'A'\frac{\partial^2 w}{\partial t^2} + \tilde{E}I'\frac{\partial^4 w}{\partial x^4} - \tilde{G}A'\frac{\partial^2 w}{\partial x^2} - \rho'I'\frac{\partial^4 w}{\partial x^2 \partial t^2} = 0, \tag{36}$$

in which $w$ is the deflection, $A'$ and $I'$ are the cross-sectional area and the cross-sectional moment of inertia, respectively. Compared with the classic *Rayleigh* beam equation, a $\partial^2 w / \partial x^2$ term is used to consider axial force. There is a traveling wave solution for Eq. (36), and the corresponding wave velocity can be expressed as

$$V_S = \sqrt{\frac{\tilde{G}A' + \tilde{E}I'k^2}{\rho'A' + \rho'I'k^2}} \approx \sqrt{\frac{\tilde{G}}{\rho'}} + k\sqrt{\frac{\tilde{E}I'}{\rho'A'}}, \tag{37}$$

since $A' \gg I'k^2$ for $k \in [0, \pi/L]$.

Fig. 6 shows $V_S$ in the rubber cord discussed in Sec. 4.1, but in some deficient pre-stress states. Both of the velocities obtained from Eq. (37) and from the numerical analysis are expressed as a function of $kL$, and yield an excellent agreement with each other. At the stress-free state, $V_S$ is nonlinearly related to $k$, and the rubber cord behaves like a beam. As the stress ($\tilde{G}$) increases, $\sqrt{\tilde{G}/\rho'}$ in Eq. (37) will quickly exceed and be much larger than $k\sqrt{\tilde{E}I'/\rho'A'}$, so that dispersion gradually weakens. As indicated in Fig. 6, $V_S$ appears as a horizontal line when the axial stress exceeds 0.2 MPa, in which case the soft fiber behaves like a string and the velocity degenerates to Eq. (22).

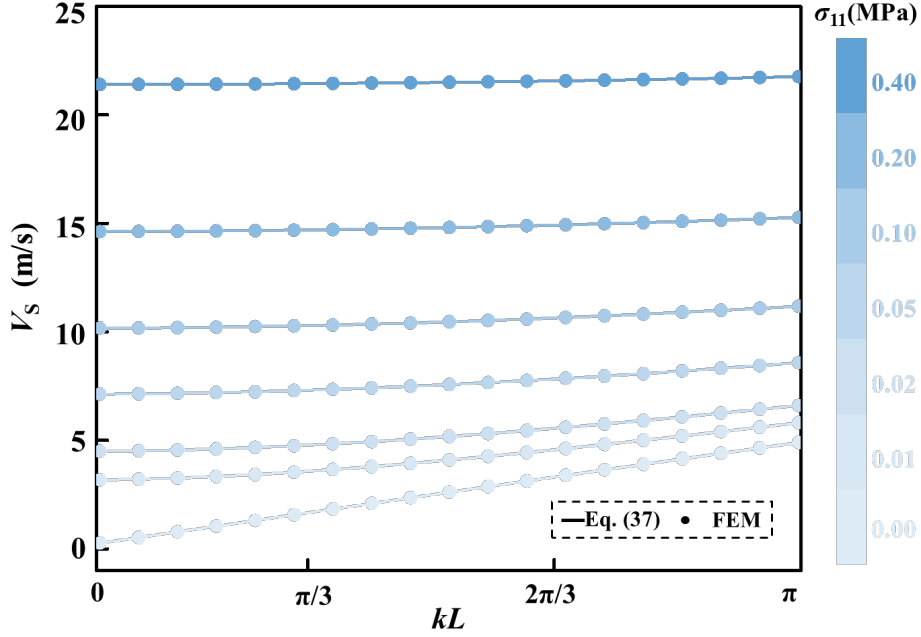

FIG. 6. S-wave velocity in pre-stretched soft fiber versus $kL$ at an extremely low stress level. As $\sigma_{11}$ increases, the dispersion gradually weakens, so that the velocity becomes independent of wavenumber $k$.

## 5. Conclusions

In summary, we have investigated linear elastic wave propagation in a finitely pre-stretched soft fiber. The governing equations have been derived based on the small-on-large theory. The formulas of $V_\mathrm{P}$ and $V_\mathrm{S}$ have been proposed in the material and stress perspectives, respectively, which have been validated by numerical simulations and experimental data on spider silk. In the material perspective, the velocities can be uniquely obtained once the constitutive model and fiber elongation are determined, which is beneficial once applied to the design and analysis of wave propagation in fiber-based complex structures. In the stress perspective, the velocities are expressed by engineering stress, elongation, and initial mass density, which is conducive to experimental measurement-related applications.

The influences of material constitution, compressibility, and pre-stress on the wave propagation have been systematically studied. With increasing pre-stress, $V_\mathrm{P}$ changes along with the concavity of the stress-strain curve, while the S-counterpart increases regardless of any constitutive model. For a soft fiber with a specific SEF, $V_\mathrm{P}$ and $V_\mathrm{S}$ may be consistent at a certain stress level, which can be anticipated in some applications such as wave mode manipulation. Moreover, for both P- and S-waves, the variation of the velocities has exhibited more significantly in a compressible fiber than that in an incompressible one, and regardless of compressibility, the increasing rate in velocities that result from the change in elongation is approximately identical.

For miniscule pre-stress, we've proposed a modified formula for $V_S$ based on the *Rayleigh* beam theory, which clearly captures the nature of dispersion of the S-wave and shows the competition between "string vibration" and "beam vibration." This finding provides a reliable theoretical basis for high-precision measurement of elastic wave velocities in soft fibers at low-stress levels.

We hope this investigation may lead to an improved understanding of the elastodynamics of soft fibers with relevant precision in mechanical characterizations. Also, we hope this work opens a promising route for lightweight, tunable wave manipulation devices.

**Acknowledgement**

This work was supported by the National Natural Science Foundation of China (Grant No.11602294).

**Appendix A: Proof of** $\tilde{E} = J^{-1} F_{11}^2 (dS_{11}/dF_{11})$

Through a similar procedure as Eq. (10) to (16), the tangent modulus which is the slope of the engineering stress–elongation curve can be derived as

$$\hat{E} = \frac{dS_{11}}{dF_{11}} = A_{1111} - \frac{A_{1122} A_{2211}}{A_{2222}}. \tag{A.1}$$

Thus, the effective Young's modulus can be obtained by pushing forward all the components of $A_{ijkl}$ appear in Eq. (A.1) into the current configuration, i.e.,

$$\tilde{E} = \frac{F_{11}^2}{J} \hat{E} = \frac{F_{11}^2}{J} \frac{dS_{11}}{dF_{11}}. \tag{A.2}$$

**Appendix B: Proof of** $\tilde{G} = \sigma_{11}$

We consider a general form of compressible SEF

$$W = W_b(J) + W_h(I_1, I_2), \tag{B.1}$$

in which $W$ is decomposed into a bulk term $W_b$ and a hybrid term $W_h$. In this fashion, the *Cauchy* stress in the current configuration

$$\sigma_{ij} = J^{-1} F_{i\alpha} \frac{\partial W}{\partial F_{j\alpha}} \tag{B.2}$$

can be expressed as

$$\sigma_{ij} = \frac{\partial W_b}{\partial J} \delta_{ij} + 2J \frac{\partial W_h}{\partial I_2} \delta_{ij} + 2J^{-1} F_{is} F_{js} \frac{\partial W_h}{\partial I_1}. \tag{B.3}$$

In the case of uniaxial tension, i.e., Eq. (10), we have

$$\sigma_{11} = \frac{\partial W_b}{\partial J} + 2J \frac{\partial W_h}{\partial I_2} + 2J^{-1} \frac{\partial W_h}{\partial I_1} \left( F_{11}^2 + F_{12}^2 \right). \tag{B.4}$$

As the upper and lower boundaries of the fiber are unconstrained, we have

$$\sigma_{22} = \frac{\partial W_b}{\partial J} + 2J \frac{\partial W_h}{\partial I_2} + 2J^{-1} \frac{\partial W_h}{\partial I_1} \left( F_{21}^2 + F_{22}^2 \right) = 0, \tag{B.5}$$

which leads to

$$\frac{\partial W_b}{\partial J} + 2J \frac{\partial W_h}{\partial I_2} = -2J^{-1} \frac{\partial W_h}{\partial I_1} \left( F_{21}^2 + F_{22}^2 \right). \tag{B.6}$$

Substituting Eq. (B.6) into Eq. (B.4), we have:

$$\sigma_{11} = 2J^{-1} \frac{\partial W_h}{\partial I_1} \left( F_{11}^{\ 2} + F_{12}^{\ 2} - F_{21}^{\ 2} - F_{22}^{\ 2} \right). \tag{B.7}$$

On the other hand, after a forward push of $A_{\alpha j \beta l}$, we have:

$$\begin{aligned}
A'_{ijkl} &= J^{-1} F_{i\alpha} F_{k\beta} A_{\alpha j \beta l} \\
&= J \frac{\partial^2 W_b}{\partial J^2} \delta_{ij} \delta_{kl} + 4 J^3 \frac{\partial^2 W_h}{\partial I_2^{\ 2}} \delta_{ij} \delta_{kl} + 2J \frac{\partial W_h}{\partial I_2} \delta_{ij} \delta_{kl} + 2 J^{-1} F_{i\beta} F_{k\beta} \frac{\partial W_h}{\partial I_1} \delta_{jl} \\
&\quad + 4 J^{-1} F_{i\alpha} F_{k\beta} F_{l\beta} F_{j\alpha} \frac{\partial^2 W_h}{\partial I_1^{\ 2}} + \left( 2 \frac{\partial W_h}{\partial I_2} + J^{-1} \frac{\partial W_b}{\partial J} \right) \left( F_{ij} F_{kl} - F_{il} F_{kj} \right),
\end{aligned} \tag{B.8}$$

in which the components related to the effective shear modulus are:

$$\begin{aligned}
A'_{1212} &= 4 J^{-1} F_{2\beta} F_{2\alpha} F_{1\alpha} F_{1\beta} \frac{\partial^2 W_h}{\partial I_1^{\ 2}} + 2 J^{-1} F_{1\beta} F_{1\beta} \frac{\partial W_h}{\partial I_1}, \\
A'_{1221} &= -\left( 2J \frac{\partial W_h}{\partial I_2} + \frac{\partial W_b}{\partial J} \right) + 4 J^{-1} F_{1\beta} F_{2\alpha} F_{1\alpha} F_{2\beta} \frac{\partial^2 W_h}{\partial I_1^{\ 2}}, \\
A'_{2112} &= -\left( 2J \frac{\partial W_h}{\partial I_2} + \frac{\partial W_b}{\partial J} \right) + 4 J^{-1} F_{2\beta} F_{1\alpha} F_{2\alpha} F_{1\beta} \frac{\partial^2 W_h}{\partial I_1^{\ 2}}, \\
A'_{2121} &= 4 J^{-1} F_{1\beta} F_{1\alpha} F_{2\alpha} F_{2\beta} \frac{\partial^2 W_h}{\partial I_1^{\ 2}} + 2 J^{-1} F_{2\beta} F_{2\beta} \frac{\partial W_h}{\partial I_1}.
\end{aligned} \tag{B.9}$$

Therefore, in combining Eq. (16), (B.8) and (B.9), we have

$$\begin{aligned}
\tilde{G} &= A'_{1212} - \frac{A'_{1221} A'_{2112}}{A'_{2121}} \\
&= 2 J^{-1} \frac{\partial W_h}{\partial I_1} \left( F_{11}^{\ 2} + F_{12}^{\ 2} - F_{22}^{\ 2} - F_{21}^{\ 2} \right) \\
&= \sigma_{11}.
\end{aligned} \tag{B.10}$$